\documentclass[fleqn,usenatbib]{mnras}

\usepackage[T1]{fontenc}

\usepackage{graphicx}
\usepackage{longtable}
\usepackage{booktabs}
\usepackage{array}
\usepackage{supertabular}
\usepackage{graphicx,multicol}
\usepackage{subfig}
\usepackage{multirow}
\usepackage{setspace}
\usepackage{float}
\usepackage{tablefootnote}
\usepackage{adjustbox}
\usepackage{hyperref}

\usepackage[normalem]{ulem} 

\usepackage{txfonts}

\usepackage{amssymb}	

\newcommand{\placetabone}{

\begin{table*}
\caption{Properties of TNG50 Fornax-like clusters and of their satellite galaxies at $z=0$.}
\label{tab:Fornax-analogues}
\centering
\resizebox{\textwidth}{!}{
\begin{tabular}{c c c c c c c}
\hline\hline
Halo ID & $R_{200c}$ $[kpc]$ & $M_{200c}$ $[10^{13}\:M_{\odot}]$ & \# of satellites & \# discs at infall & \# discs at $z=0$ & \# discs formed after infall \\
(1) & (2) & (3) & (4) & (5) & (6) & (7) \\
\hline

   2 & $846.2$ & $6.46$ & 49 & 15 & 2 & 0  \\
   
   3 & $692.4$ & $3.54$ & 32 & 10 & 2 & 1  \\
    
   4 & $671.2$ & $3.23$ & 29 & 6 & 2 & 0  \\
   
   6 & $688.7$ & $3.48$ & 31 & 11 & 1 & 0  \\
   
   7 & $680.3$ & $3.36$ & 17 & 4 & 0 & 0  \\
    
   8 & $637.6$ & $2.77$ & 20 & 5 & 0 & 0  \\
   
   9 & $675.7$ & $3.29$ & 26 & 10 & 3 & 1  \\
   
   10 & $609.7$ & $2.42$ & 12 & 4 & 1 & 0  \\
   
   11 & $661.7$ & $3.09$ & 23 & 5 & 2 & 0  \\
   
   13 & $611.1$ & $2.43$ & 21 & 7 & 3 & 1  \\
    
    \hline
\end{tabular}
}
\footnotesize
\textbf{Notes.} (1) Halo i.e. FoF ID; (2) virial radius; (3) virial mass; (4) number of satellites, i.e. of galaxies within the virial radius $R_{200c}$ and with $M_{\star}\geq3\times10^{8}$ $M_{\odot}$ at $z=0$, excluding the centrals; (5) number of galaxies (among the ones in column 4) that had stellar disc morphology at the time of accretion; (6) number of galaxies (among the ones in column 4) that have survived as discs since accretion; (7) number of galaxies (among the ones in column 4) that are discy at $z=0$ but were not discy at the time of accretion. \\
\end{table*}

}

\newcommand{\placetabtwo}{

\begin{table}
\caption{Fundamental observational properties of the three edge-on disc galaxies of Fornax cluster.}
\label{tab:obsdisc-properties}
\centering
\begin{adjustbox}{width=0.5\textwidth}
\begin{tabular}{c c c c c c}
\hline\hline
Galaxy & $m_{r}$ $[mag]$ & $M_{\star}$ $[10^{10}\;M_{\odot}]$ & $R_{e}$ $[kpc]$ & $d$ $[Mpc]$ & $d_{clus}$ $[Mpc]$ \\
(1) & (2) & (3) & (4) & (5) & (6) \\
\hline \vspace{2mm}

   FCC\,153 & $11.70$ & $0.76$ & $1.15$ & $21.73^{+1.44}_{-1.66}$ & 1.827 \\ \vspace{2mm}
   
   FCC\,170 & $10.99$ & $2.25$ & $1.45$ & $19.57^{+1.45}_{-1.38}$ & 0.407 \\ \vspace{2mm}
    
   FCC\,177 & $11.80$ & $0.85$ & $1.47$ & $17.80^{+1.88}_{-1.95}$ & 2.166 \\
   
    \hline
\end{tabular}
\end{adjustbox}
\footnotesize
\textbf{Notes.} (1) Galaxy name (of Fornax Cluster Catalog); (2) apparent magnitude in SDSS $r$-band from \citet{iodice2019} Table 1; (3) total stellar mass from \citet{iodice2019}; (4) effective radius in kpc from \citet{ferguson1989} in the B band; (5) line-of-sight distance towards the galaxy from PNe distance measurements from \citet{Spriggs2021}. The distance of the center of the Fornax from us is estimated at 19.8 Mpc \citep{Spriggs2021}; distance with respect to the central galaxy of Fornax NGC 1399 as estimated in \citet{Spriggs2021} using PNe.\\
\end{table}

}

\newcommand{\placefigProjectedHalo}{
\begin{figure*}
    \centering
    \includegraphics[scale=0.31]{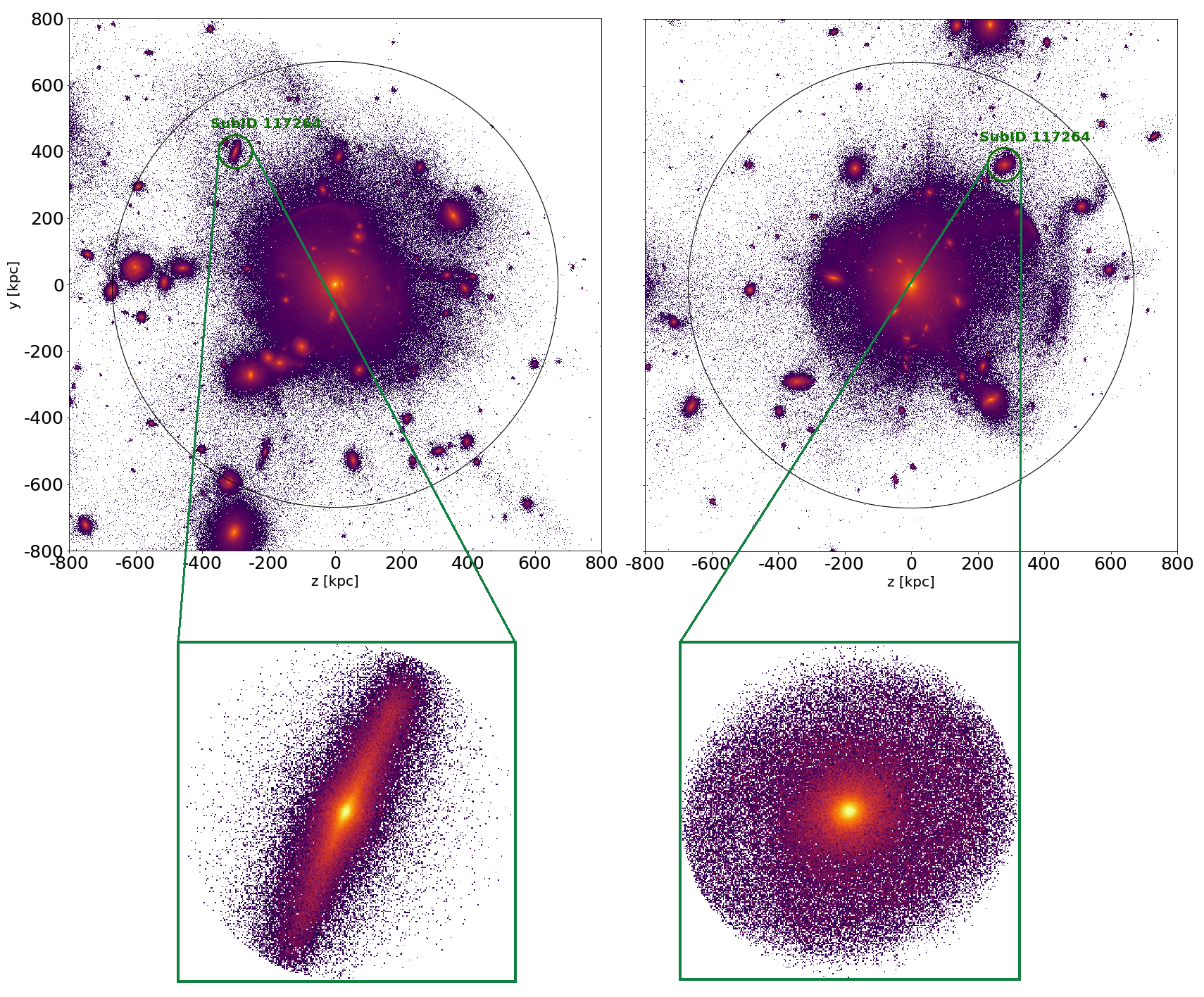}
    \caption{Two different random projections of Halo ID 3 from TNG50 at $z=0$ along with a zoom-in view of galaxy with Subhalo ID 117264 as it looks in these two random projections. The black circle on top panels show the virial radius $R_{200c}$ of the halo. This example illustrates how the same galaxy can be seen edge-on in the random LoS on the left whereas it looks in a non edge-on view on the right LoS. For the two random projections, we show the stellar particles of Subhalo ID 117264 within $2R_{\star,1/2}$. We do not include any interloper in this two random LoS but just the members of the FoF halo according to the Fornax3D-like selection described in Section~\ref{sec:projections}.}
    \label{fig:proj_halo}
\end{figure*}
}

\newcommand{\placefigHaloesFigs}{
\begin{figure}
    \centering
    \includegraphics[width=0.9\columnwidth, trim={3cm, 10cm, 3cm, 0cm}]{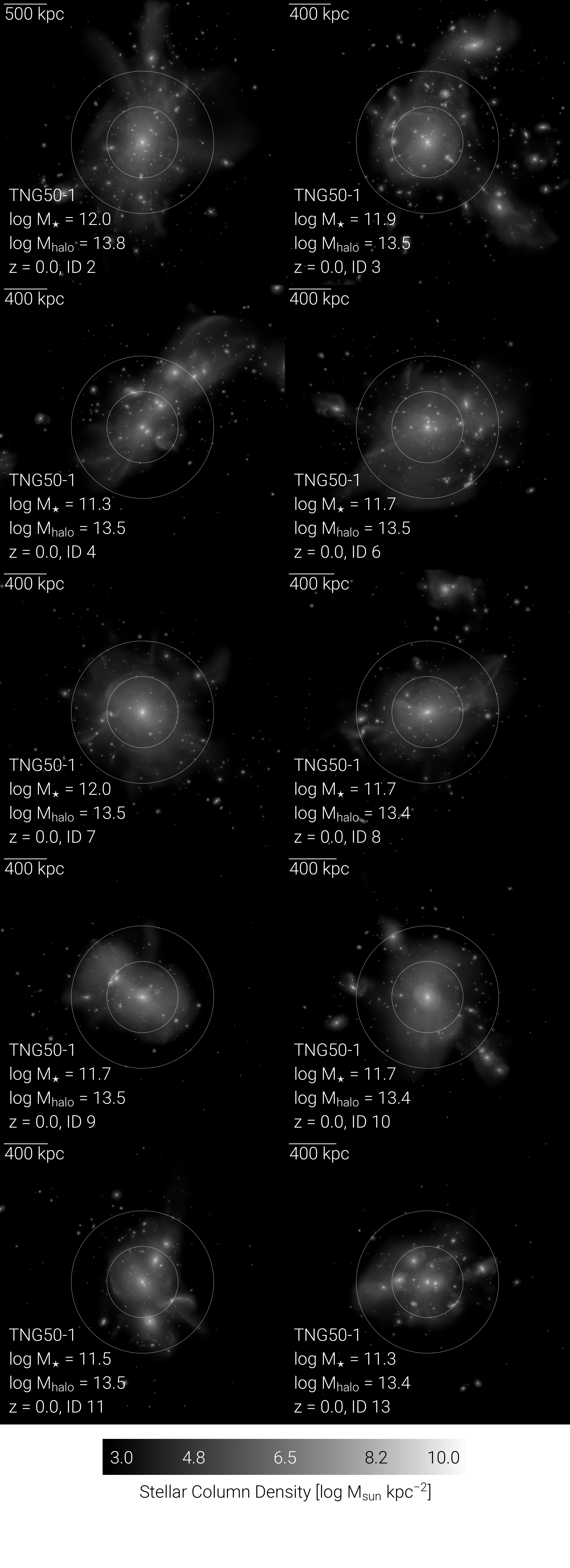}
    \caption{Stellar mass density of the 10 Fornax cluster analogues in TNG50 at $z=0$ (more properties in Tab.~\ref{tab:Fornax-analogues}). The haloes are arranged starting with Halo 2 on upper left panel and Halo 13 on bottom right panel. The haloes are shown in a random projection. The white larger circles show the virial radius $R_{200c}$ of each halo and the smaller show half of the virial radius.}
    \label{fig:haloes}
\end{figure}
}

\newcommand{\placefigMassDist}{
\begin{figure*}
    \centering
    \includegraphics[scale=0.15]{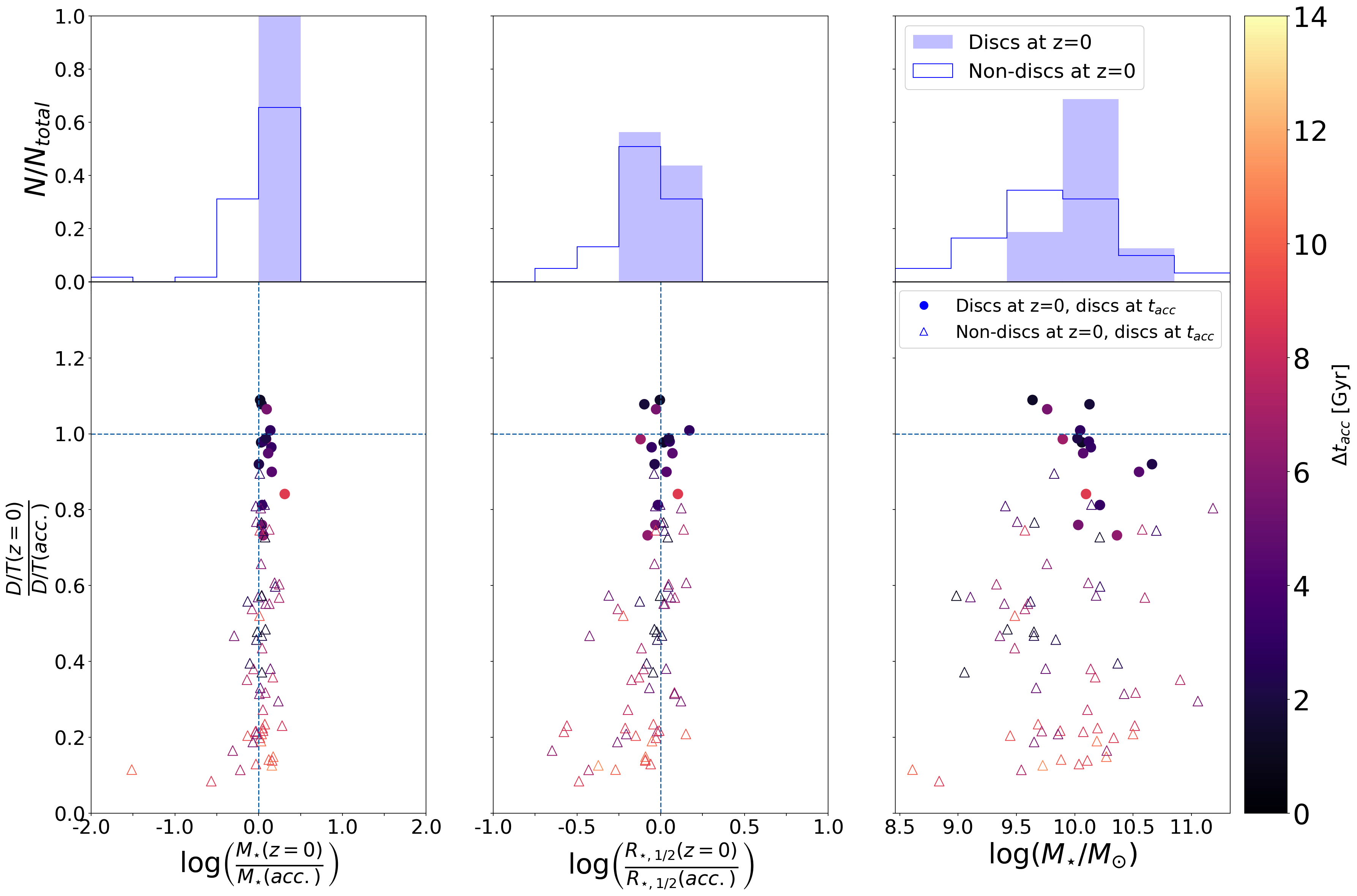}
    \caption{Changes in stellar morphology vs changes in stellar mass (left column), vs variations in stellar half-mass radius (middle column), and vs total stellar mass at present time (right panel) for cluster galaxies in TNG50 Fornax analogues. These are selected at $z=0$ to be FoF members within the virial radius of their host and to have $M_{\star}\geq3\times10^{8}$ $M_{\odot}$ (see Section~\ref{sec:selection} and Tab.~\ref{tab:Fornax-analogues}). Changes are evaluated between the time of accretion and $z=0$. Filled circles represent all galaxies that remain discy since accretion and non-filled triangles show all galaxies that have been accreted as discs and have lost their discy shape after accretion. The stellar mass comparison of first panel is considered within twice stellar half mass radius whereas the stellar mass in right panel is the total mass of all stellar particles. The color of the markers show the time since the satellite has been accreted by its final Fornax host analogue. Histograms on top show the binned mass variation, binned stellar size variation and total stellar mass distribution for discs at $z=0$ (filled bins) and non-discs at $z=0$ (non-filled bins). We do not include in this plot three TNG50 satellites that turn into discs after accretion.}
    \label{fig:circ_vs_mass-rad_dist}
\end{figure*}
}

\newcommand{\placefigTNGFornax}{
\begin{figure}
    \centering
    \includegraphics[scale=0.25]{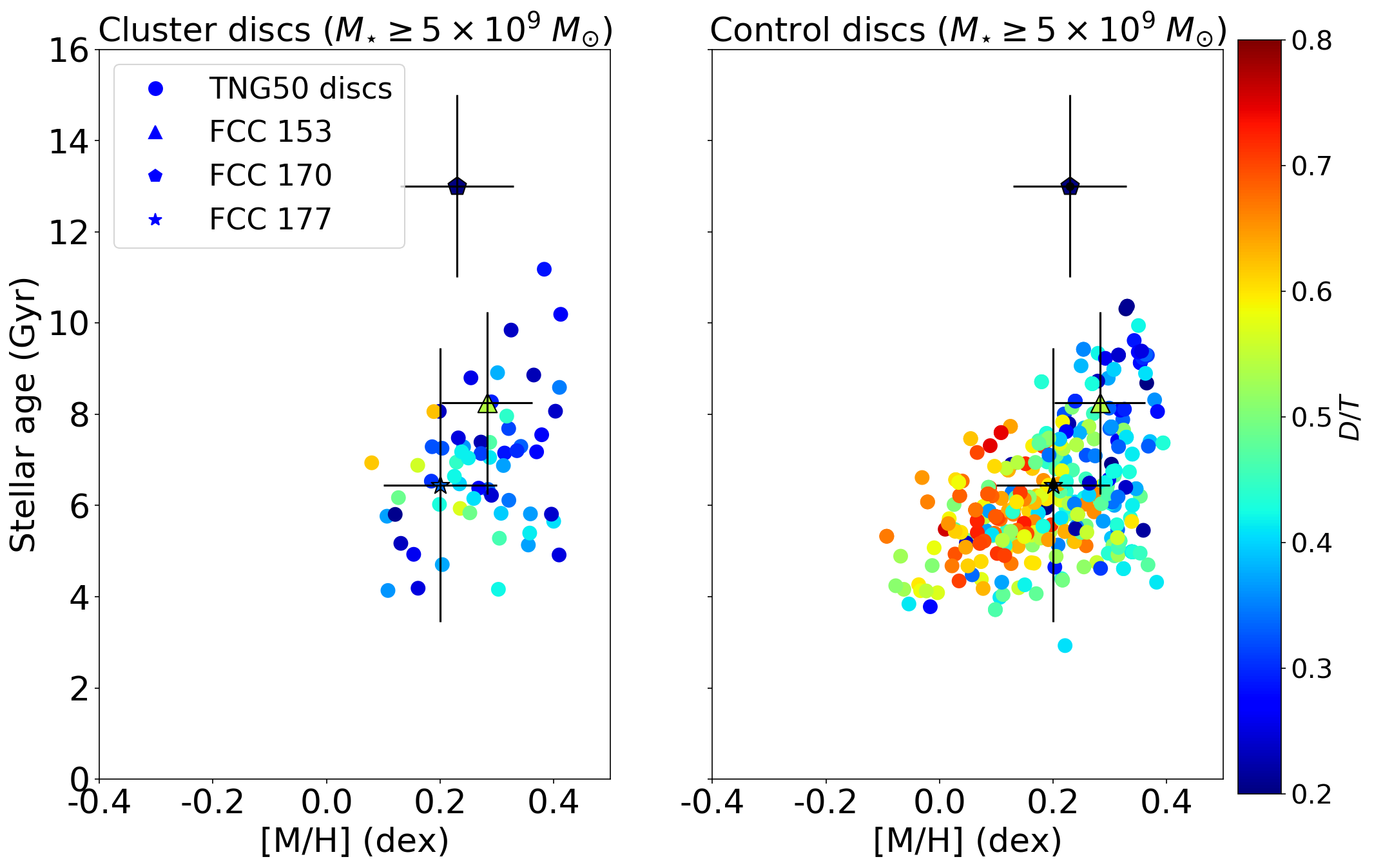}
    \caption{Mass-weighted stellar age vs metallicity near the equatorial plane for galaxies with $z=0$ values of $D/T\geq 0.2$ and $M_{\star}\geq5\times10^{9}$ $M_{\odot}$ within TNG50 Fornax-like clusters (left panel color points) and for their field counterpart matched in stellar mass (right panel color points), compared to corresponding values of the three Fornax edge-on discs FCC\,153, FCC\,170, and FCC\,177 (triangles). Disc-over-total fraction for the three Fornax edge-on galaxies come from \textcolor{blue}{Ding et al. in prep} and are on average 40\% for FCC~153 and FCC~170 and around 15\% for FCC~170. Here, stellar population properties of observed galaxies come from full spectral fitting, whereas directly from the simulation particle properties for TNG50 galaxies. Stellar discs as old and metal rich as observed in FCC\,170 cannot be found in the TNG50 simulation.}
    \label{fig:metal-age}
\end{figure}
}

\newcommand{\placefigDistances}{
\begin{figure}
    \centering
    \includegraphics[scale=0.21]{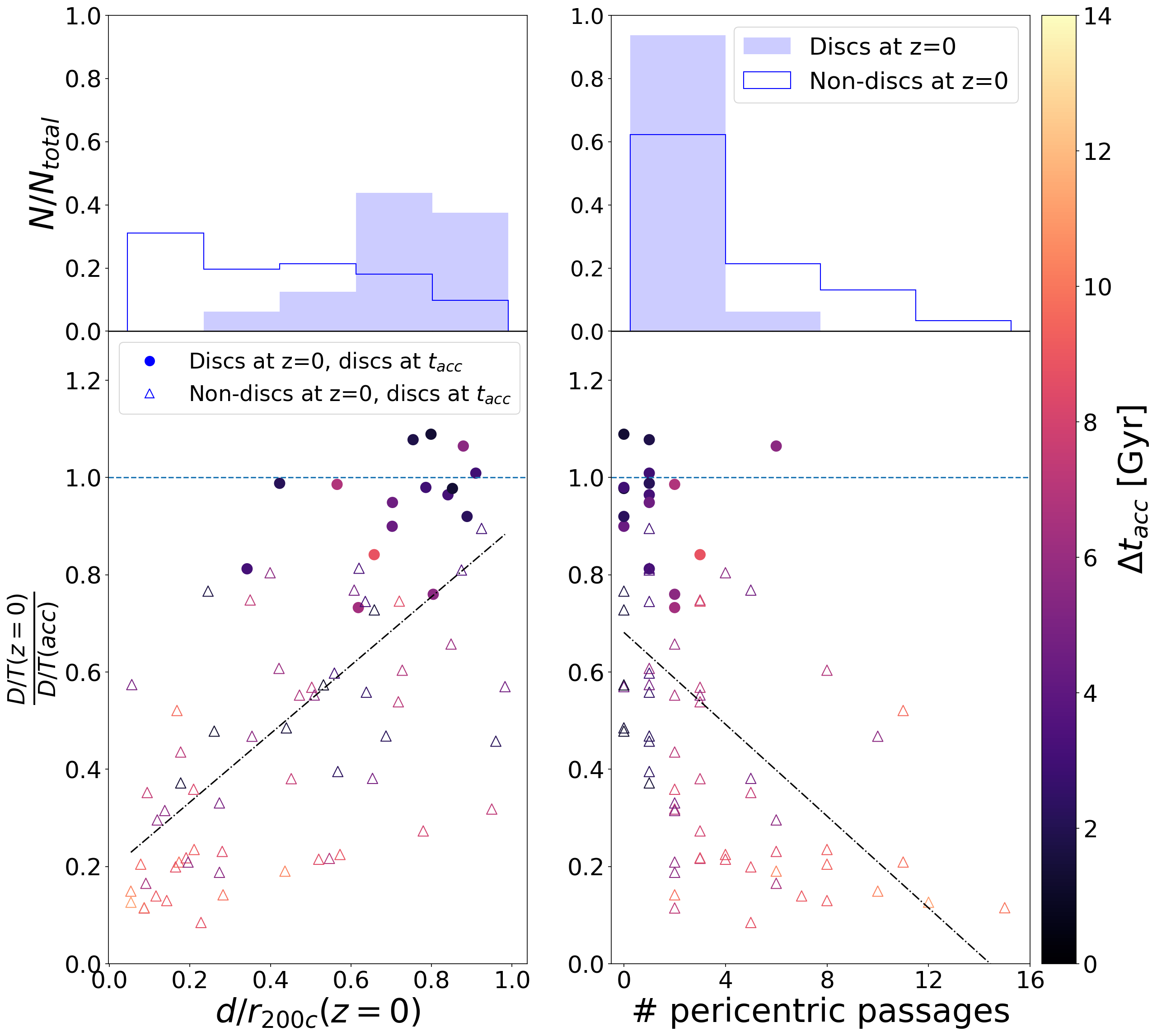}
    \caption{Morphological changes of galaxies in TNG50 Fornax analogues vs cluster-centric distance (left panel) and vs number of past pericentric passages (right panel), selected as in Fig.~\ref{fig:circ_vs_mass-rad_dist} and Tab.~\ref{tab:Fornax-analogues}. Galaxies that remain discy since accretion are plotted with circles, whereas galaxies that entered the cluster being discy and lost their disciness are shown with non-filled triangles. Time since each galaxy was accreted by their final host halo is represented by the colorbar. Dashed lines show the linear correlations for both distributions. We have not included the three galaxies that formed discs after accretion in this figure.}
    \label{fig:distances}
\end{figure}
}

\newcommand{\placefigAngles}{
\begin{figure*}
    \centering
    \includegraphics[scale=0.43]{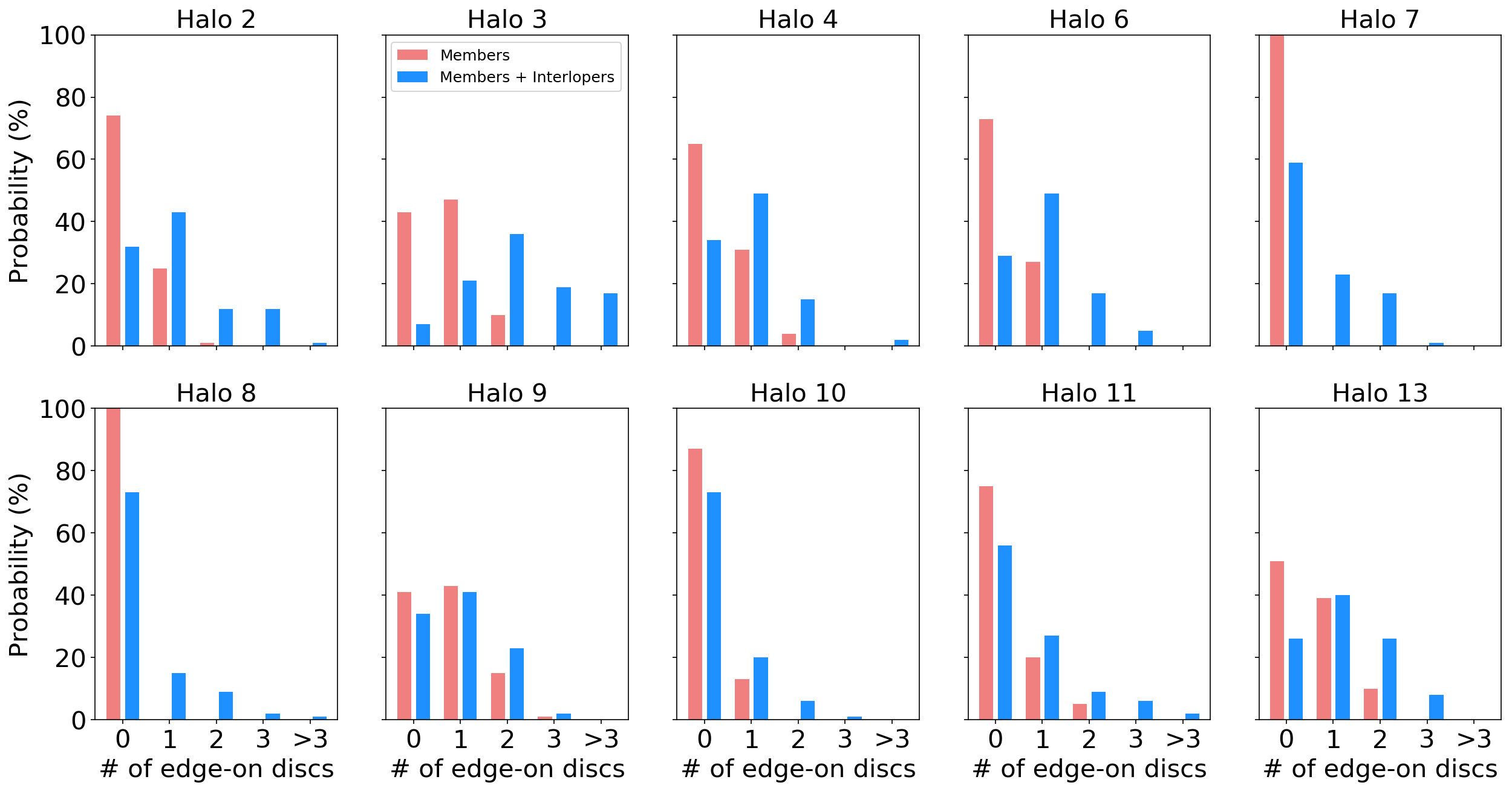}
    \caption{Probability distribution for the number of TNG50 disc galaxies brighter than 15 r-band magnitudes that would appear edge-on (i.e. with an inclination $\geq 80$ degrees) around each of our 10 Fornax analogues, when  looking at them from 100 random line of sights. We consider either cluster members only (red histograms, see Fornax3D-like selection described in Section~\ref{sec:projections}) or both galaxy members and interlopers (blue histograms).}
    \label{fig:angles}
\end{figure*}
}

\newcommand{\placefigEdgeOn}{
\begin{figure}
    \centering
    \includegraphics[scale=0.38]{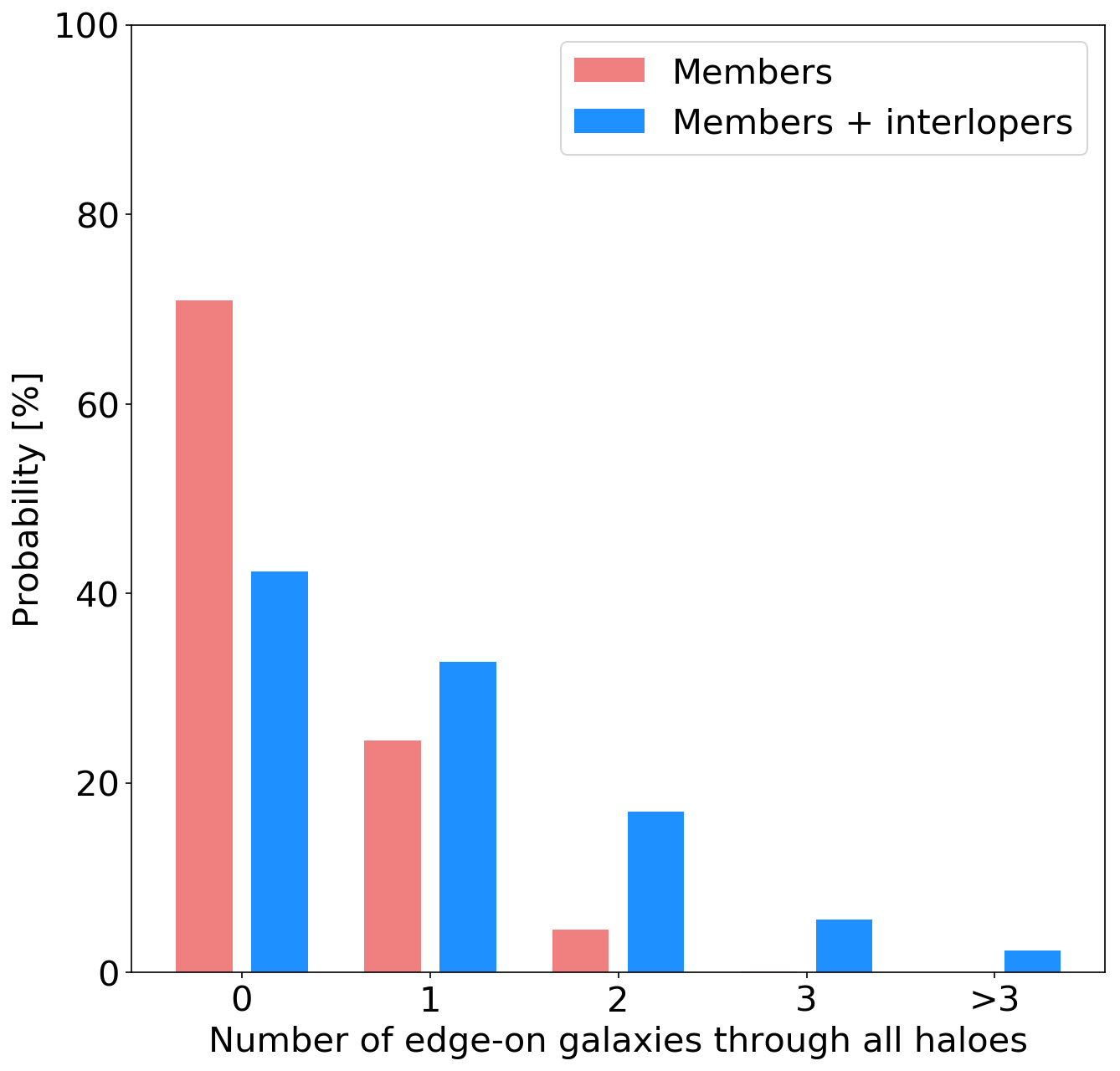}
    \caption{Same as Fig.~\ref{fig:angles} but combining all 10 Fornax-like haloes in TNG50. In this case, the probabilities have been estimated by counting the number of edge-on disc galaxies per LoS for each halo.}
    \label{fig:edge-on_discs}
\end{figure}
}

\title[The survival of stellar discs in Fornax-like environments]{The survival of stellar discs in Fornax-like environments, from TNG50 to real galaxies}

\author[P. M. Gal\'an-de Anta et al.]{Pablo M. Gal\'an-de Anta,$^{1,2}$\thanks{E-mail: pgalandeanta01@qub.ac.uk}
M. Sarzi,$^{2}$
A. Pillepich,$^{3}$
Y. Ding,$^{4,5}$
L. Zhu,$^{4}$
L. Coccato,$^{6}$
E. M. Corsini,$^{7,8}$
\newauthor
K. Fahrion,$^{9}$
J. Falcón-Barroso,$^{10,11}$
D. A. Gadotti,$^{6}$
E. Iodice,$^{12}$
M. Lyubenova,$^{6}$
I. Mart\'in-Navarro,$^{10,11}$
\newauthor
R. M. McDermid,$^{13}$
F. Pinna,$^{3}$
G. van de Ven$^{14}$
and P. T. de Zeeuw$^{15,16}$
\\
$^{1}$Astrophysics Research centre, School of Mathematics and Physics, Queen's University Belfast, Belfast BT7 INN, UK\\
$^{2}$Armagh Observatory and Planetarium, College Hill, Armagh, BT61 9DG, UK\\
$^{3}$Max-Planck-Institut für Astronomie, Königstuhl 17, D-69117 Heidelberg, Germany\\
$^{4}$Shanghai Astronomical Observatory, Chinese Academy of Sciences, 80 Nandan Road, Shanghai 200030, China\\
$^{5}$School of Astronomy and Space Sciences, University of Chinese Academy of Sciences, No. 19A Yuquan Road, Beijing 100049, China\\
$^{6}$European Southern Observatory, Karl-Schwarzschild-Straße 2, 85748, Garching bei München, Germany\\
$^{7}$Dipartimento di Fisica e Astronomia `G. Galilei', Università di Padova, Vicolo dell'Osservatorio 3, 35122, Padova, Italy\\
$^{8}$INAF-Osservatorio Astronomico di Padova, vicolo dell'Osservatorio 5, 35122, Padova, Italy\\
$^{9}$European Space Agency, European Space Exploration and Research Centre, Keplerlaan 1, 2201 AZ, Noordwijk, The Netherlands\\
$^{10}$Instituto de Astrofísica de Canarias, Vía Láctea s/n, 38205, La Laguna, Tenerife, Spain\\
$^{11}$Depto. Astrofísica, Universidad de La Laguna, Calle Astrofísico Francisco Sánchez s/n, 38206, La Laguna, Tenerife, Spain\\
$^{12}$INAF-Astronomical Observatory of Capodimonte, via Moiariello 16, I-80131
Napoli, Italy\\
$^{13}$Department of Physics and Astronomy, Macquarie University, Sydney, NSW, 2109, Australia\\
$^{14}$Department of Astrophysics, University of Vienna, Türkenschanzstrasse 17, 1180, Vienna, Austria\\
$^{15}$Sterrewacht Leiden, Leiden University, Postbus 9513, 2300 RA, Leiden, The Netherlands\\
$^{16}$Max-Planck-Institut für Extraterrestrische Physik, Giessenbachstraße, 85741, Garching bei München, Germany
}

\date{Accepted XXX. Received YYY; in original form ZZZ}

\pubyear{2022}

\begin{document}
\label{firstpage}
\pagerange{\pageref{firstpage}--\pageref{lastpage}}
\maketitle

\begin{abstract}

We study the evolution of kinematically-defined stellar discs in 10 Fornax-like clusters identified in the TNG50 run from the IllustrisTNG suite of cosmological simulations. We considered disc galaxies with present-day stellar mass $M_{\star}\geq3\times10^{8}$ $M_{\odot}$ and follow their evolution since first entering their host cluster. Very few stellar discs survive since falling in such dense environments, ranging from 40\% surviving to all being disrupted. Such survival rates are consistent with what reported earlier for the two more massive, Virgo-like clusters in TNG50. In absolute terms, however, the low number of present-day disc galaxies in Fornax-like clusters could be at odds with the presence of three edge-on disc galaxies in the central regions of the actual Fornax cluster, as delineated by the Fornax3D survey. When looking at the Fornax analogues from random directions and with the same selection function of Fornax3D, the probability of finding three edge-on disc galaxies in any one Fornax-like cluster in TNG50 is rather low, albeit not impossible. We also compared the stellar-population properties near the equatorial plane derived from integral-field spectroscopy for the three edge-ons in Fornax to similar line-of-sight integrated values for present-day disc galaxies in TNG50. For one of these, the very old and metal-rich stellar population of its disc cannot be matched by any the disc galaxies in TNG50, including objects in the field. We discuss possible interpretations of these findings, while pointing to future studies on passive cluster spirals as a way to further test state-of-the-art cosmological simulations.

\end{abstract}




\section{Introduction}

During the last decades, many efforts have been made to understand the role of galactic environment in the formation and evolution of galaxies. Many studies have indeed shown that galaxies in dense environments such as clusters follow distinct formation and evolutionary paths from those galaxies in the field, resulting in greater fractions of early-type galaxies, redder colours and lower rates of star formation compared to field galaxies \citep{dressler1980,balogh2004,hogg2004,Kauffmann2004,Alpaslan2015}.
As galaxies can and do transit from low to high-density environments over time, a number of mechanisms have been proposed to account for both morphological changes and the quenching of star formation. These include galaxy mergers \citep{barnes1996,makino1997,angulo2009} and interactions \citep{moore1996,moore1998} for the former and ram-pressure stripping \citep{gunn1972,abadi1999,yun2019} of atomic gas by the cluster X ray-emitting gas for the latter, with the activity of central supermassive black holes also playing a part in preventing the cooling of the hot cluster atmospheres \citep[e.g.][]{Fabian2011}. The relative importance of these and other mechanisms across the entire range of cluster environments and galaxy mass is not well established, however.

In this respect, focusing on the formation and evolution of the dynamically-cold and thin stellar discs in galaxies proves to be very informative, since indeed stellar discs are the natural result of star-formation activity and conversely can be significantly disrupted by mergers and interactions.
From a theoretical standpoint, \citet[][hereafter J20]{Joshi2020} followed for the first time in fully-cosmological hydrodynamical galaxy simulations the formation and evolution of disc satellite galaxies to understand the impact of cluster environments on galaxy morphology.
More specifically, they made use of the TNG50 and TNG100 runs \citep{marinacci2018,naiman2018,pillepich2018b,springel2018,nelson2018a,pillepich2019,nelson2019} of the IllustrisTNG cosmological simulation, to track the evolution of 
disc stellar structures in Virgo-like clusters since they entered such host environments.
By monitoring the fraction of stars in high-circularity orbits, they traced changes in the morphology of cluster disc galaxies, showing that the loss of cold stellar orbits comes with star formation quenching, gas removal, dark matter loss, and decrease in stellar mass and is due to tidal shocking at pericenters.

Environmental effects on satellite evolution are generally seen, in both observations and simulations, to vary between clusters of different total mass \citep{Kauffmann2004,blanton2005,Alpaslan2015}, in particular as regards the relative role of gravitational and hydrodynamical processes. For instance, as pointed out by \citet{serra2017}, in less-massive but more common group-mass hosts, gravitational interactions are expected to be more common, since galaxies move at relatively lower velocities compared to the case of galaxies in high-mass, high velocity-dispersion clusters. However, fast-moving galaxies are expected to suffer more the impact of ram-pressure stripping. Thus in Fornax-like clusters we may expect stellar thin-disc morphologies to be more likely to be disrupted than in more massive Virgo- or Coma-like clusters, but we may also expect satellites to be able to hold on to their gas reservoirs and to form stars for longer times since entering the cluster.

From an observational point of view, the nearby Fornax cluster offers an ideal laboratory to investigate the environmental impact on galaxy evolution and in particular on their disc components. This cluster environment is the second most massive cluster within 20 Mpc (after Virgo cluster) with a virial mass and radius of $M_{200c}=7\times10^{13}$ $M_{\odot}$ and $R_{200c}=0.7$ Mpc \citep{drinkwater2001}, respectively. The distance towards Fornax cluster has been established to be at 19.8 Mpc according to the last estimates \citep{Blakeslee2009,Blakeslee2010,Spriggs2021}. The Fornax cluster has indeed been extensively studied \citep[]{Hodge1978,Loewenstein1993,Graham1998,drinkwater2001,Karick2003}, including the use of integral-field spectroscopic data such as the early work of \citet{Scott2014} on the dynamics of early-type galaxies in this cluster or the more comprehensive magnitude-limited study of galaxies inside the virial radius of Fornax by the Fornax3D (F3D) project \citep{sarzi2018,iodice2019}. In particular, the F3D study of \citet[][hereafter P19a,b]{pinna2019FCC170,Pinna2019b} explored the kinematics and stellar populations of three edge-on disc galaxies: IC 1963 (FCC 153), NGC 1381 (FCC 170), and NGC 1380A (FCC 177). The analysis made in P19a,b were later complemented by \citet{Poci2021} who produced orbit-superposition dynamical and stellar-population models for the three edge-on Fornax galaxies that allowed to reconstruct their star-formation history. In addition, \citet{martin-navarro2021} explored the initial mass function of these and other Fornax galaxies. Such studies make these three edge-on disc galaxies particularly interesting for the purpose of testing the predicted evolution of stellar discs in the context of cosmological simulations. 

In the present work, we utilise the highest-resolution run of IllustrisTNG, TNG50 \citep{pillepich2019, nelson2019}, to study the effects that Fornax-like clusters exert on galaxies primarily dominated by a stellar disc component. This is done in the context of a state-of-the-art cosmological large-volume simulation, to test it and to establish similarities between simulated cluster satellites and the three observed Fornax edge-on discs. TNG50 is the only simulation currently available \citep[and publicly available:][]{nelson2019a} that follows the evolution of more than a few Fornax-mass like clusters while simultaneously resolving the inner structures of galaxies with baryonic mass resolution of $\lesssim 10^5\,M_{\odot}$ and spatial resolution of the order of $\sim100$ pc.

This paper is organised as follows: in Section 2 we present an overview of the used observational data along with a description of TNG50 and the followed procedure to make our analysis; Section 3 presents the analysis on the survival rates of Fornax thin-discs and its comparison with Virgo-like clusters; in Section 4 we discuss about the alignment of thin-disc galaxies with respect to the line-of-sight (LoS) in simulated Fornax clusters; in Section 5 we make a discussion about stellar populations in TNG50 disc galaxies and compare with their real counterparts in Fornax cluster; finally in Section 6 we give our conclusions.

\section{Observations and simulations}

In this section we explain the observational data used to compare with the disc galaxies found in TNG50 along with a brief summary of the IllustrisTNG simulations, and in particular its highest resolution run TNG50. We also explain the procedure followed in this study to identify Fornax analogues, disc galaxies in the analogues and how to get all physical quantities we use in the paper.

\subsection{The Fornax cluster}

We retrieve observational data of Fornax galaxies from the MUSE integral-field unit installed at Very Large Telescope (VLT) on the UT4. This data is part of the Fornax3D project as explained in \citet{sarzi2018} and \citet{iodice2019}. In particular, throughout the paper we focus on the three edge-on discs in Fornax: FCC\,153, FCC\,170, and FCC\,177 (IC\,1963, NGC\,1381, and NGC\,1380A, respectively). These targets host thin-disc components recently studied in dynamical models \citep{Poci2021}, which exhibit old metal-rich stellar populations extended towards the kpc-scale galactic disc as shown in P19a,b and more recently in \citet{martin-navarro2021}. The fundamental properties of the galaxies are tabulated on Table~\ref{tab:obsdisc-properties}.

The MUSE datacubes were obtained using the wide-field mode with a spatial sampling of $0.2\arcsec\times0.2\arcsec$ over a $1\arcmin\times1\arcmin$ field of view. For MUSE, the wavelength range is between 4650 and 9300 \AA\ with a spectral sampling of $1.25$ \AA\ pixel$^{-1}$ and with an average spectral resolution FWHM$_{\rm int}=2.8$~\AA. Each one of these three galaxies was covered by two pointings: a central pointing and an offset pointing that covers the outer disc and halo region of the targets. The datacubes are extensively described in \citet{sarzi2018} and \citet{iodice2019}.

The stellar populations of FCC\,153, FCC\,170 and FCC\,177 were explored in both P19a,b and \citet{martin-navarro2021} using two different approaches to obtain 2D maps of the stellar populations of the underlying stars along the regions inside $4R_{e}$ of the targets. In particular, we make use of the retrieved stellar population analysis made in P19a,b. In these complementary studies, they make use of the Penalized Pixel-Fitting method \citep[pPXF][]{Cappellari2004,Cappellari2017} based on a penalised likelihood approach. This routine permits to determine the line-of-sight velocity distribution parameters by using Gauss-Hermite expansion series to fit the spectra with a set of stellar population templates. In particular, the full-spectral fitting permits to extract both stellar ages and metallicities through the weights assigned to each individual stellar population model. P19a,b use the MILES stellar library based on BaSTI isochrones \citep[described in][]{Vazdekis2015} with each single stellar population corresponding to a combination of stellar ages and metallicities that can be averaged with the weights to infer these properties from each galaxy spectra.

\placetabtwo


\subsection{The TNG50 simulation}

We use the publicly-available output \citep{nelson2019a} of the cosmological magnetohydrodynamical simulation called TNG50\footnote{We focus on data available for TNG50 and not for larger volumes as we want the best resolution at short scales that can better reproduce the properties of observed galaxies than lower resolution runs.} \citep[][]{pillepich2019,nelson2019}. This is the highest resolution run of the IllustrisTNG project and encompasses a periodic volume of (35 $h^{-1}$ Mpc)$^{3}$. The other flagship runs of IllustrisTNG include TNG300, which covers a volume of (205 $h^{-1}Mpc$)$^{3}$, and TNG100 (75 $h^{-1}Mpc$)$^{3}$.

TNG50 is based on the $\Lambda$CDM cosmology with cosmological parameters from \citet[][density parameters $\Omega_{m} = 0.3089$, $\Omega_{\Lambda} = 0.6911$, $\Omega_{b} = 0.0486$, normalisation $\sigma_{8} = 0.8159$, spectral index $n_{s} = 0.9667$, and Hubble parameter  $h=0.6774$]{planck2015}. It has been run with  the moving-mesh code AREPO \citep{springel2010pur} and the IllustrisTNG galaxy formation model \citep{weinberger2016, pillepich2018a} from $z=127$ to the current epoch, with data saved at 100 snapshots  between $z\sim20$ and $z=0$.

The cubic volume in TNG50 was initialised with the same number $(2160^{3})$ of dark matter particles and gas cells, with a fraction of the latter eventually turning into stars and getting refined or derefined across time. The average mass of each component is $m_{DM}=4.5\times10^{5}\;M_{\odot}$ for dark matter particles and $m_{baryon}=8.5\times10^{4}\;M_{\odot}$ for gas cells and stellar particles. As regards of  gravitational softening lengths, DM and stellar particles use a softening scale of 288 pc, kept constant from $z=0$ to $z=1$, whereas gas cells have a typical size of $100 - 200$ pc in the star-forming regions of galaxies and use an adaptive gravitational softening length with a minimum at 74 pc \citep[Fig. 1 of][]{pillepich2019}.

\subsection{Identifying Fornax analogues and disc galaxies in TNG50}
\label{sec:selection}

In TNG50, groups of galaxies and single galaxies are identified by the friends-of-friends \citep[FoF,][]{davis1985} and SUBFIND \citep[][]{springel2001} algorithms, respectively. The FoF algorithm is run over the dark matter particles, and the baryonic content of each FoF is assigned to the nearest bound dark matter particle. We define a `host halo' as all the particles (regardless of type) contained within the virial radius $R_{200c}$, described as the radius where the mean density is equal to 200 times the critical density of a flat universe. Naturally, the virial radius has assigned the virial mass $M_{200c}$. In the following, every time we refer to a group/halo property, it is limited to the virial radius of the cluster unless differently specified.

In order to identify Fornax analogues in TNG50, we choose FoF groups according to their virial mass and virial radius and accounting for the observed properties of the Fornax cluster found by \citet{drinkwater2001}. These use the method of \citet{diaferio1999} to estimate the mass distribution of the cluster based on galaxy velocities. 

We select TNG50 Fornax analogues as those FoF haloes that satisfy the following conditions:
\begin{itemize}
    \item $R_{\rm Fornax}-0.2<R_{200c}$/Mpc$<R_{\rm Fornax}+0.2$ where $R_{\rm Fornax}=0.7$ Mpc and assuming a 30\% error in virial radius;
    \item $\log(M_{\rm Fornax})-0.5<\log(M_{200c})<\log(M_{\rm Fornax})+0.5$, with $M_{\rm Fornax}=7\times10^{13}\;M_{\odot}$.
\end{itemize}
These upper/lower limits assumption for virial radius and mass are consistent with the sample of Fornax analogues in \textcolor{blue}{Ding et al. in prep} to offer a number of Fornax-like clusters that statistically provide a sufficient number of haloes to compare with the properties of the Fornax cluster. The allowed range in virial cluster mass is more generous than the range in virial radius as we want to obtain a representative number of haloes that are less massive than Virgo-like cluster ($M_{200c}\sim 10^{14-14.6}$ $M_{\odot}$ as adopted in J20) but sufficiently numerous to achieve a good statistics.

By applying the previously mentioned criteria, a representation of the 10 TNG50 Fornax-like clusters that we find in the TNG50 volume at $z=0$ is plotted in Fig.~\ref{fig:haloes} starting at Halo 2 on the top left panel and finishing at Halo 13 on the bottom right panel. The properties of the analogues are presented on Tab.~\ref{tab:Fornax-analogues} indicating the number of satellites found at present time and the number of those satellites that were discs at accretion, disc galaxies at present time, and galaxies accreted as non-discs that eventually turn into discs. It should be noted that the initial conditions of the TNG50 simulation were chosen so that the resulting volume at $z=0$ is consistent with the average high-mass end of the dark-matter halo mass function given the adopted cosmological model \citep[see][for more details]{pillepich2019}: the number of 10 Fornax-like clusters is hence what would be expected in a typical volume of 50 comoving Mpc$^3$.

\placefigHaloesFigs

For each FoF halo, subaloes are identified as the collection of material that makes self gravitationally-bound objects. These objects are embedded within FoF haloes in a hierarchical structure if they fall within the volume enclosed within a FoF halo, but in the simulation box there can also be individual subhaloes that are not members of any host, and are thus called central or field (sub)haloes.
We identify a galaxy as any subhalo within the TNG50 box with non-zero masses of both stars and dark matter at present time $z=0$. The position of a galaxy is given by its minimum potential particle or cell, irrespective of type and its velocity is given by the mass-weighted average velocity of all its constituents. For the FoF groups, the position of the group itself is identified as the position of its central galaxy. Every property referred to a galaxy is constrained within twice the stellar half mass radius $R_{\star,1/2}$.

As done in J20, we adopt the same procedure to define a sample of cluster galaxies that belong to the Fornax analogues. Making use of the SUBFIND algorithm, we identify all galaxies that belong to each FoF Fornax-like cluster and that are found at $z=0$ within their virial radius. We exclude central galaxies as they follow rather different evolutionary pathways \citep[e.g.][]{Mistani2016, pillepich2018b, Joshi2020}. Throughout this paper, we consider only galaxies with $M_{\star}\geq 3\times10^{8}\,M_{\odot}$ at $z=0$, namely with at least roughly 3500 stellar particles each \citep{Joshi2020}. The most massive satellite galaxy in our TNG50 Fornax-like sample has a stellar mass of $3.1\times10^{11}$ $M_{\odot}$.

As regards of the identification of disc galaxies in TNG50, we first calculate the circularity of satellite galaxies \citep[as defined in][and done in J20 Eq. 1]{marinacci2014} and then adopt the morphological classification criterion defined in J20 and previously explained in \citet{genel2015}. To calculate the circularity of satellites we first align the galaxies with the galactic plane using the tensor of inertia of the stars within $2R_{\star,1/2}$ in same way as done in J20 and \citet{pulsoni2020}. Then, we take the specific angular momentum of each stellar particle perpendicular to the galactic plane of the galaxy to further compare with the maximum angular momentum of the 50 particles before and 50 particles after the particle in question arranging them by their specific binding energy.

To define the `disciness' of a galaxy we measure the fraction of stellar particles with circularities $\varepsilon>0.7$. This threshold has been proven to be reliable to show up thin-discs in simulated galaxies \citep{aumer2013}. We adopt the definition of the disc-to-total mass ratio (or circularity fraction)
\begin{equation}
    \frac{D}{T}=\frac{\sum_{i} M_{i}(\varepsilon>0.7)}{\sum_{i} M_{i}},
\end{equation}
with $M_{i}(\varepsilon>0.7)$ the mass of a \textit{i}-th stellar particle with circularity above 0.7, and $M_{i}$ the mass of the \textit{i}-th stellar particle regardless of its circularity. This formula gives the fraction of the total stellar mass that is contributed by high circularity stellar particles i.e. stellar particles in cold orbits. We call disc galaxies those with $D/T>0.4$ galaxies and non-disc galaxies those with $D/T<0.4$, as in J20.

Last, to measure how cluster environments affect the stellar structure of $z=0$ satellites, we follow the simulated galaxies back in time. For this, we use SubLink\_gal \citep{rodriguez2015} to track the baryonic content of a given galaxy's merger tree.  Given the $z=0$ satellites of the cluster sample defined above, we measure their stellar morphologies also at the \textit{time of accretion or infall}. We adopt the definition of accretion time of J20 as the last snapshot before a satellite becomes part of its final FoF host. For this, we seek for the time at which the two merger-tree branches of host halo and satellite galaxy do not belong to the same central i.e. FoF halo. That time is defined as the sought accretion time of the satellite.

\section{Survival of thin-discs in Fornax analogues}
\label{sec:survival}

As shown in J20, disc galaxies in Virgo-like clusters ($M_{200c}=10^{14.0-14.6} \, M_{\odot}$) tend to lose their disc-like stellar morphology after been accreted by their $z=0$ cluster host, and more frequently than in the field. J20 quantified the typical survival rate of disc galaxies to be in the range $13-43$\% in the 2 Virgo-like clusters of TNG50 and $0-30$\% across the 14 TNG100 hosts (with an average of 5\% disc survival for TNG100 satellites with $M_{\star}\geq5\times10^{9}$ $M_{\odot}$). By comparison, for field galaxies in TNG50 the survival rate found by J20 is, on average, about 70\%. Here we quantify how these phenomena unfold in lower mass but possibly more compact hosts than Virgo.
 
As in J20, we start by presenting in Fig.~\ref{fig:circ_vs_mass-rad_dist} how the morphological variation of disciness since the time of accretion into Fornax-like hosts (quantified through D/T) relates to changes over the same time interval in stellar mass (within $2R_{\star,1/2}$, left column), changes in the stellar size (in terms of their stellar half-mass radius, middle column) and to galaxy stellar mass at $z=0$. This is shown for disc galaxies that retained their disc structure since being accreted (filled circles) and disc galaxies that since then lost their disc morphology (non-filled triangles).

The number of TNG50 satellites in the considered sample is given in Tab.~\ref{tab:Fornax-analogues}, for each TNG50 Fornax-like cluster. Tab.~\ref{tab:Fornax-analogues} shows that three satellite galaxies (in three different clusters) developed a stellar disc {\it after} accretion: these are interesting cases that may deserve a dedicated study. In the following two figures, for clarity, we omit them from the plots but comment on them as appropriate. Moreover, Tab.~\ref{tab:Fornax-analogues} shows that, already at accretion, a non-negligible fraction of satellites were already non-discy: as noted in J20, this is most certainly due to pre-processing in lower-mass hosts prior to falling into the $z=0$ Fornax-like hosts.

Fig.~\ref{fig:circ_vs_mass-rad_dist} shows a picture for the evolution of discs that is rather similar to what found by J20 (see their Figs.~6 and 8). As shown by the top histograms, on average disc galaxies that retained their disc experienced a modest increase in mass and little-to-no variation in size. On the other hand, objects that lost their disc characteristics since the time of accretion display a wider range of size variations often shrinking in size, with little to no variations in mass. The three galaxies that developed a stellar disc after accretion (not shown) instead grew both in stellar mass and size. 

In terms of present-day stellar mass, surviving disc galaxies tend to be somewhat more massive than objects that lost their discs, although these latter present a wide range of present-day mass values that also captures well the stellar mass of the three edge-on Fornax galaxies (see Tab.~\ref{tab:obsdisc-properties}). Significant disc component are notable absent below a present-day mass below $M_{\star}\sim3\times10^{9}$ $M_{\odot}$, in agreement with the findings of J20 for more massive hosts. Looking at the mass of disc galaxies at the time of accretion further indicates that objects below this mass appears could not retain their discs. J20 also found a clear correlation between the morphological variation of their disc galaxies and the distance from the centre of the cluster at present time and an anticorrelation with the number of pericentric passages. In Fig.~\ref{fig:distances} we see that the discs in Fornax analogues follow the same behaviour whereby only those objects that end up at relatively large distances from the centre and went through only a few or no pericentric passages could retain their discs. The colorbar in Figs.~\ref{fig:circ_vs_mass-rad_dist} and \ref{fig:distances} also shows the time since accretion to the cluster, and further indicates that objects that entered clusters as discs and retain this structure were generally accreted more recently than objects whose discs were destroyed. Indeed, while surviving discs have been accreted less than 4 Gyr ago on average, non-surviving discs were mostly accreted 6 Gyr ago. For completeness, we note that the three objects that formed a disc after infall (not shown) are rather close to the centre of their cluster at $z=0$ (less than $0.2$ $d/R_{200c}$) but have gone through only a few pericentric passages (less than 4 times).

Concerning the fraction of galaxies that entered our TNG50 Fornax-like clusters as discs and that retained their disc morphology, this varies from $43\%$ (halo 13, Tab.~\ref{tab:Fornax-analogues}) to all discs being disrupted (halos 7 and 8), with a median survival rate across the whole sample of $\sim$21\%. This survival rate falls in between the one found for the two Virgo-like clusters of TNG50 by J20 (13 and 43\%, respectively). On the one hand, in less massive hosts we can expect more low-velocity galaxy encounters and more pericenter passages inducing tidal shocking, which indeed is the mechanism identified by J20 for turning cold into hotter orbits. In Fornax-like clusters, Fig.~\ref{fig:distances} (right panel) indicates that indeed the number of pericentric passages is still a key element in driving disc disruption. On the other hand, in less massive hosts, satellites may be able to retain their gas reservoirs against ram pressure stripping for longer times, hence retaining their capability of keeping forming a stellar thin disc. Moreover, galaxy harassment \citep{moore1996} may also play a significant role in the survival of disc galaxies, in particular since higher velocity dispersion systems promote rapid fly-bys, leading to disc disruption in low-mass galaxies. Consequently, in lower mass clusters galaxy harassment could be less probable to occur then leading to a higher chance for discs to survive. Given the limited high-mass cluster sample size in TNG50, we are not able to identify a host-mass trend, also because, as shown by J20 (their Appendix A), the host-to-host variation is connected to the distribution of the satellites' accretion times, which can be substantially different for different hosts i.e. different hosts' assembly histories.

On the one hand, in less massive hosts we can expect more low-velocity galaxy encounters and more effective tidal shocking at pericenter passage, which indeed  is the mechanism identified by J20 for turning cold into hotter orbits, albeit for Virgo-like clusters

\placefigMassDist
\placefigDistances
\placetabone

In terms of actual numbers of present-day discs in the Fornax-analogues, Tab.~\ref{tab:Fornax-analogues} shows that only a handful of disc satellites can be found in any given cluster, at most 4. Whether these numbers are consistent with the presence in the actual Fornax cluster of the three well-studied disc galaxies FCC\,153, FCC\,170 and FCC\,177 \citep[e.g.,][]{pinna2019FCC170,Pinna2019b}, it needs to be determined. In fact, we need to keep in mind that the observed galaxies are nearly perfectly edge-on objects. 
This suggests that there could be several more Fornax galaxies with substantial embedded stellar discs (as revealed through careful dynamical modelling, see e.g., the case of FCC\,167 shown in \citet{sarzi2018} and also in \textcolor{blue}{Ding et al. in prep}). Next, we hence assess the probability of finding disc galaxies in and around Fornax-like clusters in TNG50 when looking at them from random lines of sight.

\section{Galaxies alignment}
\label{sec:projections}

To check if the number of present-day disc galaxies in the TNG50 Fornax analogues is consistent with the number of edge-on galaxies observed in the real Fornax cluster, we need to evaluate the number and orientation of TNG50 disc galaxies that we would observe when looking at those TNG50 Fornax analogues from different directions. 

We proceed in two complementary ways. Firstly, we measure probabilities from TNG50 only considering member cluster galaxies within a certain cylinder along the line of sight as done in Fornax observations and as described below. Secondly, we relax the assumption that the three edge-on disc galaxies of Tab.~\ref{tab:obsdisc-properties} are indeed Fornax members (see 5th column of Tab.~\ref{tab:obsdisc-properties} and discussion below) and re-evaluate probabilities including galaxy interlopers from TNG50: in this context, an interloping galaxy would be any galaxy that does not satisfy the Fornax-like membership criteria but falls within the LoS as a foreground/background galaxy.

We start this exercise by drawing 100 random LoS from which we can observe the entire TNG50 simulation as well as each of the TNG50 Fornax analogues listed in Tab.~\ref{tab:Fornax-analogues}. It should be noticed that, from now on, we drop the satellite selection of e.g. Tab.~\ref{tab:Fornax-analogues} and consider all galaxies in the TNG50 simulation box above our mass limit and irrespective of whether they are part of any FoF halo.
For each LoS, we apply a corresponding set of rotations to the original positions and peculiar velocities of every galaxy in the TNG50 simulation so that the LoS corresponds to the new x-axis. The view of a galaxy through a random LoS is illustrated in Fig.~\ref{fig:proj_halo} where we plot two different random projections of Halo ID 3: one galaxy appears edge-on for one of the random LoS whereas it looks like a non edge-on and possibly a non-discy galaxy in the other LoS.
We then consider each Fornax cluster analogue at a time, shifting the position of all galaxies in the simulation so that the central galaxies of the TNG50 cluster is at the distance of the actual Fornax cluster.
We take 19.8 Mpc as the distance towards the Fornax cluster obtained from last estimations of \citet{Spriggs2021} using the planetary nebulae luminosity function.

To select Fornax3D-like member galaxies, we apply the same selection function for projected and LoS distances from the cluster centre that we would find in Fornax galaxies. We compute SDSS r-band apparent magnitudes at the new LoS distance of each TNG50 galaxy, as well as its peculiar velocities as done in the Fornax3D survey \citep{sarzi2018,iodice2019}. 
In this new reference frame and with such LoS velocities and apparent magnitudes, we firstly apply a magnitude limit as in the observations, i.e. a magnitude cut at a total r-band magnitude of 15, as done for the Fornax3D sample. We hence proceed to select only TNG50 galaxies within an aperture of 0.9 Mpc radius\footnote{We take an aperture of $0.7+0.2$ Mpc which corresponds to the virial radius of Fornax cluster as obtained by \citet{drinkwater2001} plus a 30\% uncertainty in the virial radius. This aperture is consistent with $R_{200c}$ of our 10 Fornax analogues as shown in Tab.~\ref{tab:Fornax-analogues}.} in the project yz-plane and centred at the position of the central galaxy of each Fornax analogue. Finally, we also exclude objects with excessively high or low LoS velocities compared to that of the cluster, with differences in excess of twice the velocity dispersion of the Fornax cluster analogue under consideration. These criteria impose an average distance along LoS of $\sim3$ Mpc in front and behind the central galaxy of each cluster, which is consistent with distance uncertainties as measured for Fornax galaxies \citep{Blakeslee2009,Blakeslee2010,Spriggs2021}.

Once we have selected objects around our TNG50 Fornax analogues, we focus only on those presenting stellar disc morphology (based on their $D/T$ values) and derive their inclination with respect to the LoS. 
For this we calculate the tensor of inertia of disc galaxies along the LoS (see Section~\ref{sec:selection}). Using the principal axes of the tensor of inertia, we can inverse the rotation to recover the random LoS of the galaxy. In particular, rotating from random axes to the principal axes of the inertia tensor corresponds to a transformation defined by three angles around three axes $R_{z}(\phi)R_{y}(\theta)R_{x}(\psi)$ (here $R_{i}(\alpha)$ defining the rotation matrix of angle $\alpha$ around $i$-th axis). We note that as our LoS for convenience is defined along the x-axis, the second rotation of angle $\theta$ around the new y-axis is the one which aligns the galactic plane of a disc galaxy with our LoS. We define edge-on disc galaxies those with an inclination angle greater than 80 degrees (corresponding to $|\,\theta\,| \leq10$ degrees).

\placefigProjectedHalo

\placefigAngles

Fig.~\ref{fig:angles} summarises the result of this exercise, showing for each of our TNG50 Fornax-analogues how often we find a given number (including none) of disc objects that appear edge-on. Fig.~\ref{fig:angles} also breaks down such a frequency of disc edge-on galaxies depending on whether these are truly cluster members (red histograms) or irrespective of this requirement and thus including also interlopers (blue histograms).
As expected from the low number of disc satellites found in any given Fornax-analogue (at most 4, Tab.~\ref{tab:Fornax-analogues}), it is rare that these objects present sufficiently similar orientations as to then appear simultaneously edge-on. Considering cluster members only, Fig.~\ref{fig:angles} indeed shows that only in one TNG50 Fornax-analogue (Halo 9) we can find three edge-on disc galaxies as observed in the real Fornax cluster, and only for one of our 100 LoS. Put it in an another way, in the volume covered by the TNG50 simulation there is only a 0.1\% probability to find three edge-on disc galaxies truly belonging to any one Fornax-like cluster\footnote{These probabilities rely on a small number of 10 hosts, that may not necessarily accurately capture the characteristic of the real Fornax cluster. In this work, we do observe each simulated cluster from 100 different LoS: however, the cluster content does not vary and may have an under-abundance of disc galaxies by construction.}, as shown in Fig.~\ref{fig:edge-on_discs}. Including interlopers considerably increases the number of disc galaxies appearing to be members of Fornax-like clusters and thus also the chances of several of them appearing discy and edge-on. Indeed, in this case there is a $\sim\!8\%$ chance of finding at least 3 edge-on discs in any given cluster, although we note that one single Fornax-analogue (Halo 3) contributes largely to this result.

Looking back at the real Fornax cluster, distance measurements alone cannot confirm whether all the three edge-on galaxies of Tab.~\ref{tab:obsdisc-properties} actually reside within the virial radius. Surface-brightess fluctuations measurements \citep{Blakeslee2009,Blakeslee2010} suggests FCC\,170 could reside just behind the clusters, whereas estimates based on the planetary nebulae luminosity function \citep{Spriggs2021} or on the combination of various measurements (i.e. from the Cosmicflow-3 catalogue of \citealt{Tully2016}) allow for all three objects to be in it.
On the other hand, both the projected position of the three edge-on galaxies in the Fornax cluster as well as their place in phase-space suggests that at least two of them, FCC\,170 and FCC\,177, have long since fallen into the cluster and belong to the prominent north-south clump of galaxies near the cluster centre and which aligns with the cosmic filament connecting to the Fornax-Eridanus large-scale structure \citep{iodice2019,Nasonova2011}.

Distances aside, the detailed orbit-superposition dynamical modelling for the Fornax edge-on galaxies \citep[][\textcolor{blue}{Ding et al. in prep}]{Poci2021} shows that FCC\,170 actually sports a relatively smaller stellar disc in mass compared to FCC\,153 and FCC\,177. Indeed, whereas the discs of FCC\,153 and FCC\,177 still contribute on average around 40\% of the stellar mass within 2$R_e$, in FCC~170 only about 15\% of the stellar mass within the same radius is found in dynamically-cold orbits. 

Repeating the previous calculations for finding edge-on disc galaxies but now with lower-mass discs contributing 20\% of the total stellar mass, we can factor in these considerations and compute the overall probability of finding across all Fornax-cluster analogues in TNG50 an edge-on galaxy like FCC~170 alongside two others with more massive discs like FCC~153 and FCC~177. This turns out to be 5\% if all three objects are truly members of the clusters, as per definition adopted in Fornax3D or 24\% if allowing for one interloping edge-on galaxy with D/T $\ge0.4$.

\placefigEdgeOn

Overall, these numbers suggest that the Fornax cluster may have a rather special structure or a rather specific assembly history that are not commonly represented by the 10 Fornax-like clusters we have identified in TNG50 and that would be more frequently captured by cosmological simulations with a larger volume than TNG50. There is evidence that galaxy spin correlates with the large scale structure, hence, the alignment of galaxies in clusters is highly dependent on their accretion history \citep{Wang2018}. If the three edge-on galaxies in Fornax were accreted from the same filament, their orientation in the cluster will not be, in any case, random. Consequently there might be a possibility for them to be edge-on that would only be observable in a larger cosmological volume.

Finally, a possible alternative interpretation of the numbers put forward by the analysis above is that there simply may be a dearth of galaxies with significant stellar disc components in Fornax-like clusters in the TNG50 simulation. This could be due to numerical effects related to the numerical resolution choices of the simulation or to a non fully-realistic rendering within the IllustrisTNG model of how high-density environments affect galaxy evolution and stellar galaxy morphology, or a combination thereof. In relation to the former, a possible reason for a lack of stellar discs among cluster satellites in TNG50 could be numerical heating \citep{Ludlow2019,Ludlow2021}. Caused by the presence of particles of different mass in the simulations, this numerical effect has been shown with gravity-only simulations to be a consequence of the equipartition theorem, which causes more massive particles to partially transfer their orbital energy to the lighter ones: for example, energy may be transferred from dark-matter particles to stellar particles, which in turn could result in the consecutively expansion or heating of thin, dynamically-cold stellar disc structures. Whether this applies to the IllustrisTNG AREPO-based simulations in general and to TNG50 satellites in particular remains to be demonstrated, as it remains unclear how the arguments above apply to the case of hydrodynamical simulations i.e. to galaxies that, a priori, include also gas. In fact, TNG50 galaxies throughout the simulation volume (i.e. irrespective of their central vs. satellite status) have been shown to exhibit also very thin stellar discs, e.g. with $\lesssim200-300$ pc stellar disc heights, and to have stellar structures qualitatively consistent with SDSS-based observational results \citep{pillepich2019}. Moreover, the TNG50 cluster satellites have been shown to exhibit quenched fractions that are in the ball park of observations \citep{donnari2021}, possibly lower than observed in the considered mass ranges by 10-20 percentage points: this could imply a somewhat larger gas mass content in TNG50 satellites than in reality, which in turn would work against the numerical heating scenario described above. Still, if  numerical heating was responsible for artificially ``puffing-up'' what would otherwise be thin-disc components or if issues with galaxy evolution were in place in TNG50 cluster galaxies, we might also find systematic differences between the stellar population properties of such discs in TNG50 galaxies and those of real disc galaxies. This is what we set out to check in the next section.

\section{Stellar populations of disc galaxies}

We compare the stellar-population properties of TNG50 disc galaxies to those of the three Fornax edge-on galaxies as reported by \cite{pinna2019FCC170} and \cite{Pinna2019b}.
If simulated discs were significantly affected by numerical heating (as speculated above), we would expect to find a systematic contamination in the simulation's disc structures by more metal-poor and older bulge-like stellar populations, or fail altogether to find very old thin discs, as more time would have been available for numerical heating to disrupt them.

The first step to compare simulations to observations is to align all TNG50 disc galaxies in the Fornax-analogues at an edge-on angle, following the same procedure based on the diagonalisation of the tensor of inertia that was explained in Sec.~\ref{sec:projections} to compute the inclination of simulated disc galaxies. Following this, we take a thin slit-like aperture along the galactic plane with a height equal to $R_{\star,1/2}/10$ to each side of the plane and a width of $2R_{\star,1/2}$ and compute overall mass-weighted stellar age and metallicity values along the line of sight from the simulation particle values. We proceed in the same way to obtain similar integrated values for the three Fornax edge-on discs using the mass-weighted stellar age and metallicity maps from full spectral fitting published in \cite{pinna2019FCC170} and \cite{Pinna2019b}. Note that we disregard here whether simulated galaxies would actually be observed edge-on in this instance, as we aim at assessing as broadly as possible the possible impact of numerical heating. 

The left panel of Fig.~\ref{fig:metal-age} shows how the integrated stellar age and metallicity values near the equatorial plane of all disc galaxies in the TNG50 Fornax analogues compare to the corresponding values for the Fornax cluster edge-on galaxies, with error bars corresponding to the standard deviation of the mass-weighted stellar age and mass-weighted metallicity along the equatorial-plane aperture. This includes both disc galaxies as defined in Sec.~\ref{sec:selection} according to J20 with $D/T>0.4$ and objects with smaller disc components down to $D/T=0.2$.
Rather than accounting for the role of interlopers through random projections, the right panel of Fig.~\ref{fig:metal-age} more simply shows the same comparison for a control sample of TNG50 disc galaxies, matched in stellar mass to the disc galaxies in the Fornax analogues. The selection of control galaxies is made by adopting the procedure in J20 considering all galaxies that are centrals at the time of accretion of each cluster galaxy and remain centrals at $z=0$. We only consider galaxies in both samples (cluster and control) with total stellar mass of $M_{\star}\geq5\times10^{9}$ $M_{\odot}$, to be more consistent with the stellar masses of FCC\,153, FCC\,170 and FCC\,177 (see Table~\ref{tab:obsdisc-properties}).
In the case of both cluster and control disc galaxies, whereas the disc stellar population properties of the Fornax edge-on FCC\,153 and FCC\,177 fall within the distribution for the stellar age and metallicity of TNG50 discs, a thin disc that is simultaneously as old and metal-rich as the one seen in the remaining Fornax edge-on FCC\,170 cannot be found across the entire TNG50 volume. In fact, it is the old stellar age, rather than the metallicity, of FCC\,170's disc that is not reproduced by TNG50: this is the case for neither Fornax analogues nor control galaxies with stellar masses similar to the Fornax edge-on objects and even when considering D/T ratio as low as 20\%.

\placefigTNGFornax

It is unclear what the implications of Fig.~\ref{fig:metal-age} are and here we hence offer considerations for future investigations. Firstly, the stellar ages of average galaxies of IllustrisTNG, in general, and of TNG50, in particular, have been shown to be consistent with SDSS observations throughout the available mass spectrum \citep[][and \textcolor{blue}{Boecker et al. in prep}, respectively]{nelson2018a}. Interestingly, \textcolor{blue}{Boecker et al. in prep} show that TNG50 reproduces also the wide spread of the galaxy-to-galaxy stellar-age variation at fixed galaxy mass that is inferred with SDSS data: however, this has not been checked for disc-like galaxies only and by focusing only on the equatorial planes of the latter. A dedicated study targeting the comparison between TNG50 to a larger observed galaxy sample may shed light. Alternatively, FCC\,170 could simply be a rather special system, which could in fact be realised in cosmological simulations with a larger volume than TNG50 (e.g. TNG100); or its characterisation may be particularly affected by the metallicity-age degeneracy that still hinders also full spectral fitting (\textcolor{blue}{Boecker et al. in prep}). Finally, ultimate test beds on state-of-the-art cosmological simulations can be found in specific classes of relatively unfrequent objects. 
In this respect, the relatively unusual population of passively-evolving spiral galaxies first noticed by \citet{vandenBergh1976} in the Virgo cluster may further help in understanding if simulations indeed lack in massive, old and metal-rich discs among cluster satellites. Such passive spirals have indeed large discs (with value for the disc over bulge ratio of 0.82 corresponding to D/T$\sim0.5$, \citealt{mahajan2020}) and stellar population as old and metal-rich as those found in lenticular galaxies \citep{pak2021} that could place tight constraints on the predictions from cosmological simulations. The TNG100 simulation has been tested against observed red spirals, with \cite{Rodriguez-Gomez2019} finding, if anything, a somewhat higher fraction of red discs in TNG100 in comparison to Pan-STARRS data. A similar analysis targeting cluster galaxies and focusing on the TNG50 and other simulations is warranted.

\section{Summary and conclusions}

We have explored the frequency and survival probability of stellar discs in galaxies in and around 10 Fornax-cluster analogues from the TNG50 cosmological magnetohydrodynamiccal simulation. We have defined disc galaxies as objects containing dynamically-identified stellar discs contributing over 40\% of the total stellar mass (within 2$R_e$) and compared the survival rates of such discs to the ones found in more massive clusters in the IllustrisTNG simulations. We have also investigated the orientation of present-day disc galaxies around TNG50 Fornax-like clusters and explored the likelihood of finding three of such systems (FCC 153, FCC 170 and FCC 177, see Tab.~\ref{tab:obsdisc-properties}) appearing edge-on as observed in the real Fornax cluster. Finally, we have compared the stellar-population properties of simulated disc galaxies near their equatorial plane when observed edge-on to those derived for those same three real objects based on MUSE integral-field observations.

In this way we have shown that:

\begin{enumerate}

    \item The chances of stellar disc structures being preserved following the entrance in their $z=0$ Fornax-like hosts is on average $\sim 21\%$ (for galaxies with $M_{\star}\geq3\times10^{8}$ $M_{\odot}$), but with a large cluster-to-cluster variation ($0-43$ \% across the 10 TNG50 systems). This is consistent to the survival rate of stellar disc structures found in more massive Virgo-like TNG50 clusters \citep{Joshi2020}. As in more massive hosts, more massive satellites at present, satellites that have spent shorter time in the cluster environment and those that have undergone fewer pericentric passages are more resilient to morphological changes.
    
    \item Fornax-like clusters in TNG50 host at most 4 galaxies with disc-like morphologies within their virial radius at $z=0$. The small number of present-day disc galaxies in Fornax-cluster analogues makes it somewhat unlikely for several of them to be simultaneously observed edge-on, with a 5\% probability for finding three edge-on cluster members in a Fornax3D-like observation while also allowing for one of them to have only a 20\% disc mass fraction as observed in one of three Fornax edge-on galaxies.
    
    \item The stellar populations of simulated TNG50 disc galaxies allow for the kind of the intermediate/young and metal-rich populations observed near the equatorial plane of the Fornax edge-on disc galaxies FCC\,153 and FCC\,177. However, there are no TNG50 galaxies with discs as old and metal-rich as in the remaining Fornax edge-on FCC\,170, regardless of whether we look in Fornax-analogue environments or in a control sample of similarly-massive central/field galaxies. 
    
\end{enumerate}

Based on the results of this analysis, we speculate that the Fornax cluster may have a rather special structure or a rather specific assembly history, which would be best or more frequently captured by cosmological simulations with a larger volume than TNG50. Alternatively or concurrently, it remains open the possibility that the thin edge-on discs found in observations may in fact be interlopers just in front or behind the Fornax cluster. Finally, the relative infrequency of disc galaxies in TNG50-Fornax analogues and the absence of very old thin discs might suggest that numerical heating may still play a part in artificially perturbing such dynamically-cold structures even in the latest of cosmological simulations; our  tests,  however,  are  unable  to confirm that hypothesis. Future studies centred on larger samples of passive spirals in massive clusters may provide strong test beds for current cosmological simulations.


\section*{Acknowledgements}

GvdV acknowledges funding from the European Research Council (ERC) under the European Union's Horizon 2020 research and innovation programme under grant agreement No 724857 (Consolidator Grant ArcheoDyn). The data used in this paper are publicly available in the ESO Science Facility Archive under programme ID 296.B-5054(A). Francesca Pinna, Ignacio Mart\'in Navarro and Jes\'us Falc\'on Barroso acknowledge support through the RAVET project by the grant PID2019-107427GB-C32 from the Spanish Ministry of Science, Innovation and Universities (MCIU), and through the IAC project TRACES which is partially supported through the state budget and the regional budget of the Consejer\'ia de Econom\'ia, Industria, Comercio y Conocimiento of the Canary Islands Autonomous Community."

\section*{Data Availability}

The data that support the analysis and results of this study are available on request from the corresponding author. The data pertaining to the TNG50 simulation are already available on the IllustrisTNG website: \url{www.tng-project.org/data}.



\bibliographystyle{mnras}
\bibliography{bibliography} 



\bsp	
\label{lastpage}
\end{document}